\newcommand{\HI}{H{\,\small I}}
\newcommand{\kms}{km s$^{-1}$}
\documentstyle[11pt,paspconf]{article}
\input psfig

\begin{document}

\title{Morphology and kinematics of \HI\  
in dust--lane elliptical galaxies}
\author{Raffaella Morganti}
\affil{Istituto di Radioastronomia, Bologna, Italy}
\author{Tom Oosterloo}
\affil{IFCTR -- CNR, Milano, Italy}
\author{Elaine M. Sadler}
\affil{School of Physics, University of Sydney, Australia}
\author{Daniela Vergani}
\affil{Istituto di Radioastronomia, Bologna, Italy}

\begin{abstract}

We present the results of the \HI\ observations of four dust-lane ellipticals. 
The \HI\ detected is regularly distributed in disks or rings.  Moreover, in
three of the four objects we have detected a very large amount of neutral
hydrogen with values of $M_{\rm \HI}/L_B$ that are at the end of the
distribution characteristic of ellipticals and similar to what typically found
in spiral galaxies.  Thus, these dust-lane ellipticals may represent the
results of major mergers as seen at a late stage of the evolution.  The
regular \HI\ distribution and kinematics has allowed to derive values of
$M/L_B \sim 20$ at distances of $\sim$5 $R_{\rm e}$ from the centre.  Evidence
for an increase of the dark matter contribution at larger radii can also be
derived from the flat rotation curves observed in some of the observed
galaxies. 

\end{abstract}


\section{Introduction}

When we are dealing with neutral gas in elliptical galaxies we know that, on
average, we will be detecting a limited number of objects and typically a
small amount of gas (Knapp et al.\ 1985).  In spite of this there are 
good reasons for studying \HI\ in these objects because, when detected, it 
gives valuable information on a number of interesting topics. 

{\bf Galaxy Evolution.} The \HI\ gas seen in ellipticals usually has 
an {\sl external origin}.  There are several pieces of evidence for
this (i.e.\ the lack of correlation between the galaxy's size and the 
amount of \HI; the frequent decoupling of stellar and kinematic axes; 
and the fact that gas usually has higher specific angular momentum 
than the stars --- for details see Knapp et al.\ 1985 and van Gorkom 1992).  
However, the story may be more complicated.  In the last few years it has 
apparent that elliptical galaxies which show ``peculiarities'' or ``fine
structures'' (Schweizer 1995) are also more likely to contain detectable 
amounts of \HI\ (Bregman et al.\ 1992, van Gorkom \& Schiminovic 1997, 
Morganti et al.\ 1997a).  This finding has been explained by van Gorkom 
\& Schiminovic (1997) either as an effect of {\sl environment} 
(i.e.\ signatures of peculiarity and the presence of neutral gas are both rare
in cluster environments, where most ellipticals are) 
or as an effect of {\sl evolutionary stage}.  In the latter case, the
``peculiarities'' would just represent a phase in the ``normal'' formation and
evolution of ellipticals (at least for the field objects).  
A question which still remains open is whether most of the gas comes from 
a series of small accretion events over time, or from major mergers. 
Recently, there have been several studies of major mergers and on the fate of
gas in these systems (Hibbard \& van Gorkom 1996; Hibbard \& Mihos 1995). 
All this makes the study of \HI\ in elliptical galaxies particularly
interesting if we want to understand the formation and evolution of this class
of objects and attempt to answer the questions: {\sl which are the ellipticals
resulting from these major mergers? can we get some clues from the \HI?} 

{\bf The dark matter problem}.  Studies of \HI\ kinematics in
elliptical galaxies can give reliable measurements of the mass--to--light ratio
($M/L$) at large radii.  Recently, Bertola et al.\ (1993) combined $M/L$ values
derived from \HI\ data and estimates for the central $M/L$ (based on optical
measurements from ionized gas disks) to show evidence for an increasing $M/L$
with radius, supporting the idea of a dark halo around ellipticals. 
However estimates of $M/L$ from \HI\ data are only available for a handful of
well-studied ellipticals (e.g.\ IC~2006 Franx et al.\ 1994; NGC~1052 van Gorkom
et al.\ 1986; NGC~4278 Raimond et al.\ 1981, Lees 1994; NGC~5266 Morganti et
al.\ 1997b).  It would be valuable to increase the number of galaxies
for which a similar analysis using \HI\ data can be done.  This requires not
only finding ellipticals with \HI, but finding those in which the \HI\ has 
settled into a regular disk. 

{\bf Fuelling of an active nucleus} It has been suggested that a connection may
exist between the gas content of an elliptical galaxy and nuclear activity 
(Gunn 1979, van Gorkom 1992): the gas could provide fuel for an active
galactic nucleus (AGN) and hence the acquisition of gas could turn on
activity in the nucleus.  {\sl However, the relation between gas content and
nuclear activity only holds in a statistical sense and many exceptions exist}. 
Thus it appears that other factors are important in determining whether an AGN
in an early type galaxy is switched on by the presence of gas, and that a more
subtle relationship exists between \HI\ content and nuclear activity. 

\section{Our \HI\ observations of dust--lane ellipticals}

With the above issues in mind, we have started {\sl a long term project} 
using the Australia Telescope Compact Array (ATCA) to
increase the number of early--type galaxies with detailed \HI\ observations. 
With this enlarged
database, the three issues mentioned above can be addressed in a more detailed
and systematic way.  Some of this work is also described by Oosterloo et
al.\ (these proceedings).  As part of this project we have observed seven 
dust--lane galaxies taken mainly from the compilation of Sadler \& Gerhard
(1985). 

Dust lanes are relatively common in elliptical galaxies.  Allowing for
inclination effects, Sadler \& Gerhard (1985) showed that up to 40\% of nearby
ellipticals can contain dust (and dust lanes).  This is also in reasonable
agreement with the fraction of ellipticals containing small amount of \HI. 
Dust--lane ellipticals are believed to be the result of accretion from mergers. 
Indeed in these galaxies the kinematics of dust and stars are often
substantially different.  However, the dust (if settled or quasi--settled) can
indicate the position of one of the principal planes of the galaxy.  
This is particularly valuable if we want to determine the intrinsic shape of
the galaxy and the mass--to--light ratio $M/L$.  The extended radio structure
observed in some dust-lane elliptical galaxies tends to be perpendicular to the
dust lane (Kotanyi \& Ekers 1979, M\"ollenhoff et al.\ 1992), implying a
possible connection between the dust lane itself and the AGN. 

Here we present the results we obtained for four dust--lane ellipticals that we
have detected in \HI.  In the three remaining objects observed (i.e.\ 
ESO 263--G48, NGC 7070A, NGC 7049) we have not detected neutral hydrogen
($M_{\rm \HI} < 2 \times 10^8 M\odot$).

\section{A brief description of the objects}

\subsection {NGC~5266}

\begin{figure}
\centerline{\psfig{figure=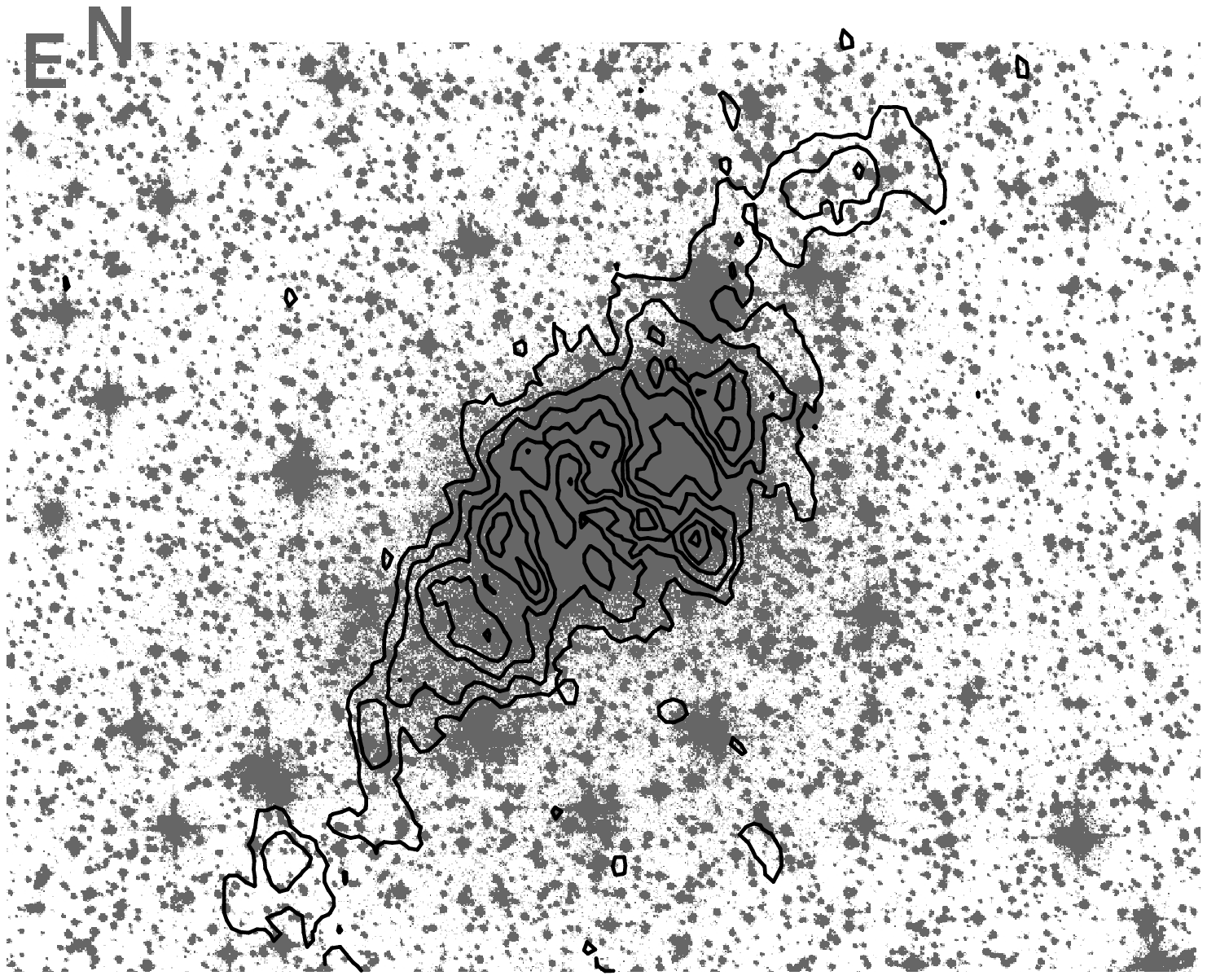,width=7cm,angle=90}
            \psfig{figure=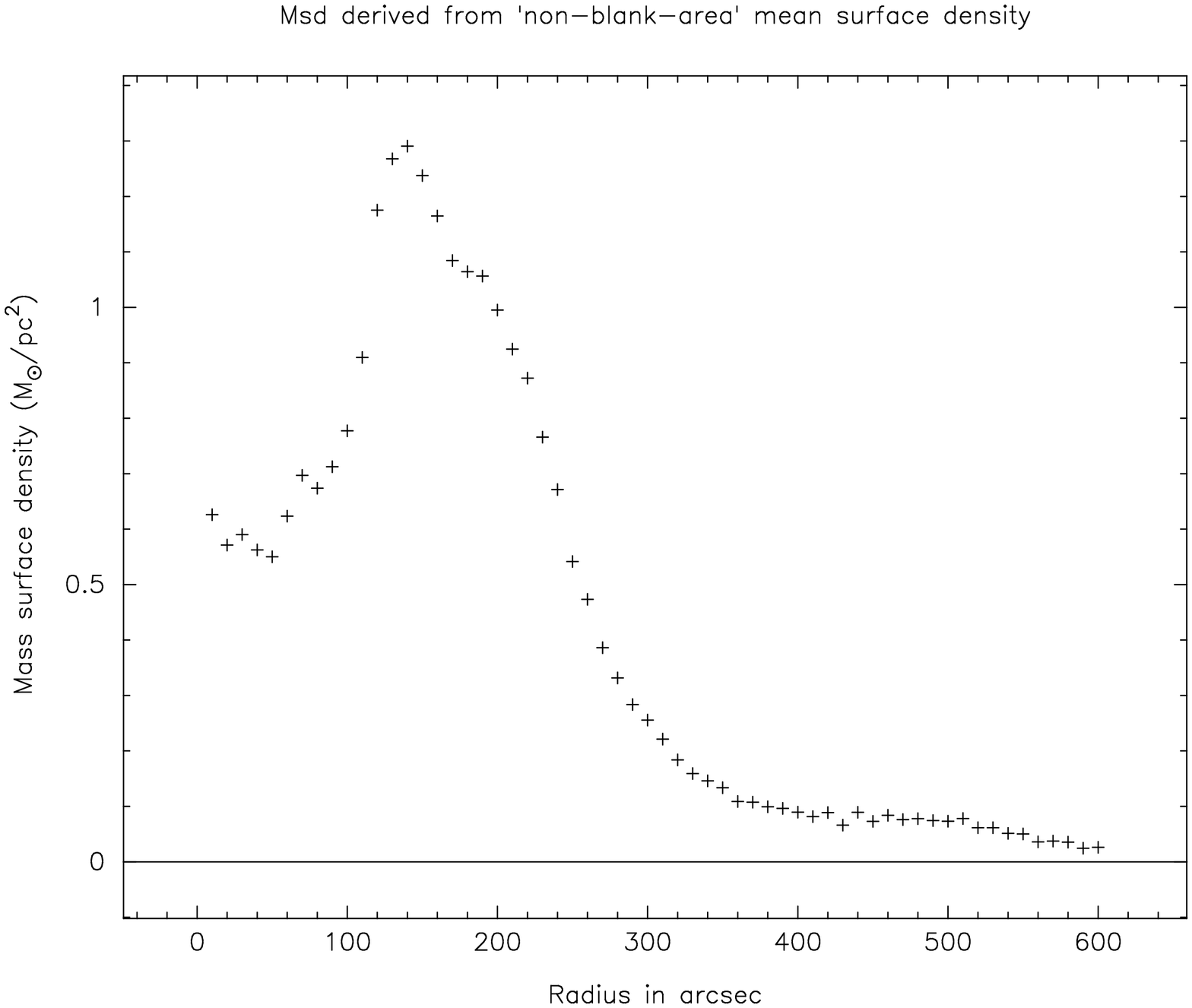,width=7cm,angle=-90}}
\caption{{\sl Left:} Total \HI\ of NGC~5266 (contours) superimposed to a deep
optical image (from D. Malin). {\sl Right:} plot of the azimuthally averaged \HI\
surface density in NGC~5266.}
 
\end{figure}

This minor--axis dust--lane elliptical galaxy contains a very large amount of
\HI\footnote{We use $H_{\circ} = 50$ km s$^{-1}$ Mpc$^{-1}$}, $\sim 2.4 \times
10^{10} M_\odot$, giving $M_{\rm \HI}/L_B \sim 0.2$.  The gas extends to $\sim$
8$^\prime$ ($\sim 140$ kpc) each side of the nucleus (see Fig.\ 1), or 8 times
the optical half--light radius $R_{\rm e}$.  Surprisingly, most of the \HI\
extends almost orthogonal to the optical dust lane (Morganti et al.\ 1997b).  A
small fraction of the \HI\ is associated with the dust lane and there are some
hints of a faint warp connecting the two structures.  The \HI\ distribution is
somewhat clumpy and asymmetric, but the overall velocity field in the inner
4$^\prime$ can be successfully modeled by assuming that the gas lies mainly in
two perpendicular planes --- in the plane of the dust lane in the central parts
and orthogonal to this in the outer regions.  Beyond the 4$^\prime$ radius, the
gas has a different structure and may be in the tidal tails, or an edge-on
ring.  The rotation velocity is at least 250 \kms\ at a radius of 4$^\prime$
and this would imply a value of $M/L_B \sim $ 8 at $\sim$4 $R_{\rm e}$.  If the
outermost \HI\ is in an edge-on ring, we estimate $M/L_B \sim$ 16 at $\sim$8
$R_{\rm e}$.  A faint continuum source is detected in the centre of NGC 5266. 

\subsection {NGC~3108}

This is an elliptical galaxy with a minor--axis dust lane and a disk of ionized
gas in the same position angle.  Luminous material is also seen in the same
plane as the dust lane (Caldwell 1984).  This galaxy is also rich in \HI: we
have detected $\sim 5 \times 10^9 M_\odot$ of neutral hydrogen (giving a value
of $M_{\rm \HI}/L_B \sim 0.14$).  The gas is distributed in a very regular ring
extended $\sim 35$ kpc each side (see Fig.\ 2).  At larger radii, the \HI\
appears to be in a warped structure.  No radio continuum emission is detected
in this object.  Assuming circular orbits and a spherical mass distribution, 
we find the $M/L_B$ to be about 22 at 5 $R_{\rm e}$. 

\begin{figure}
\centerline{\psfig{figure=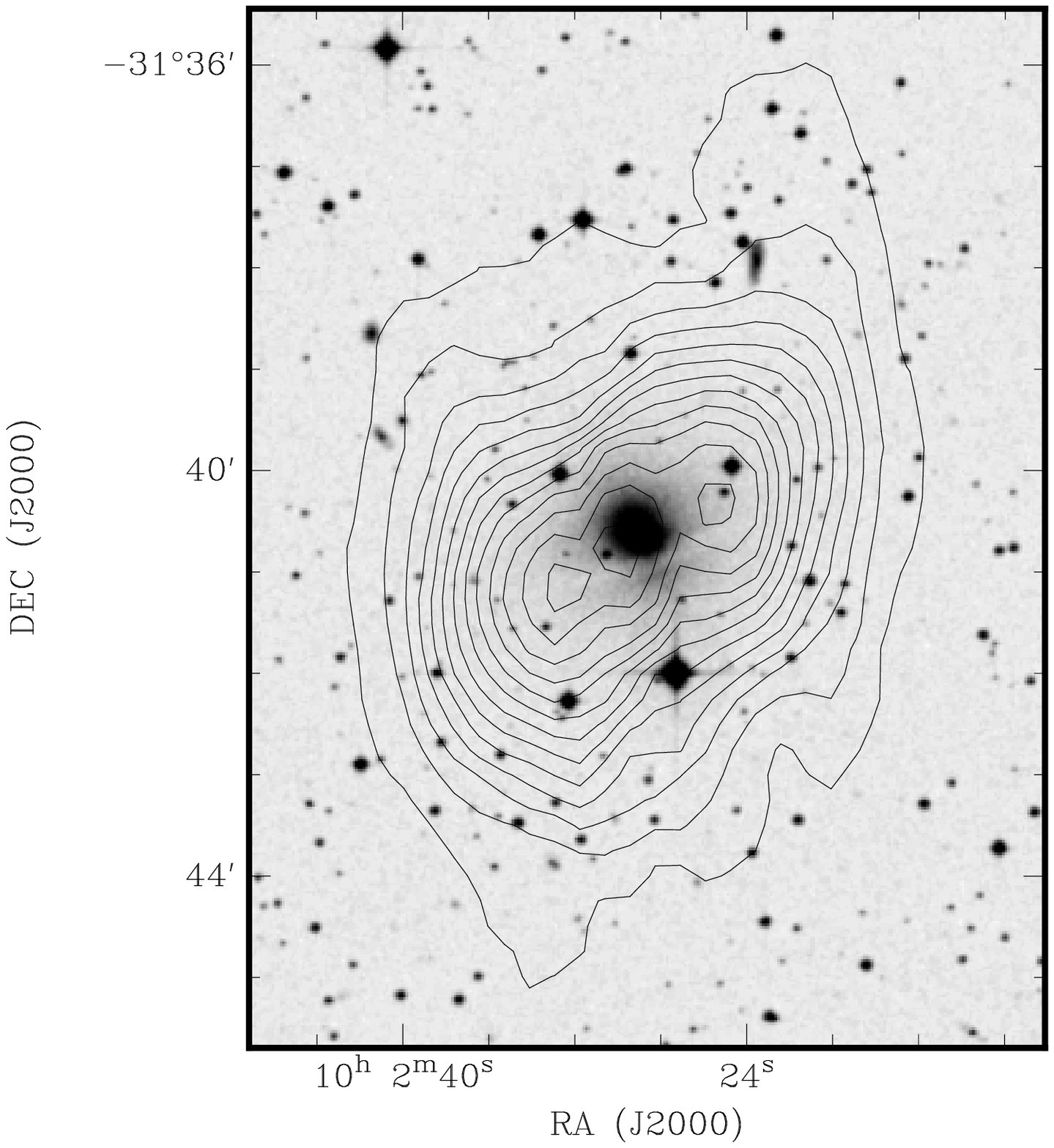,width=8cm,angle=90}
            \psfig{figure=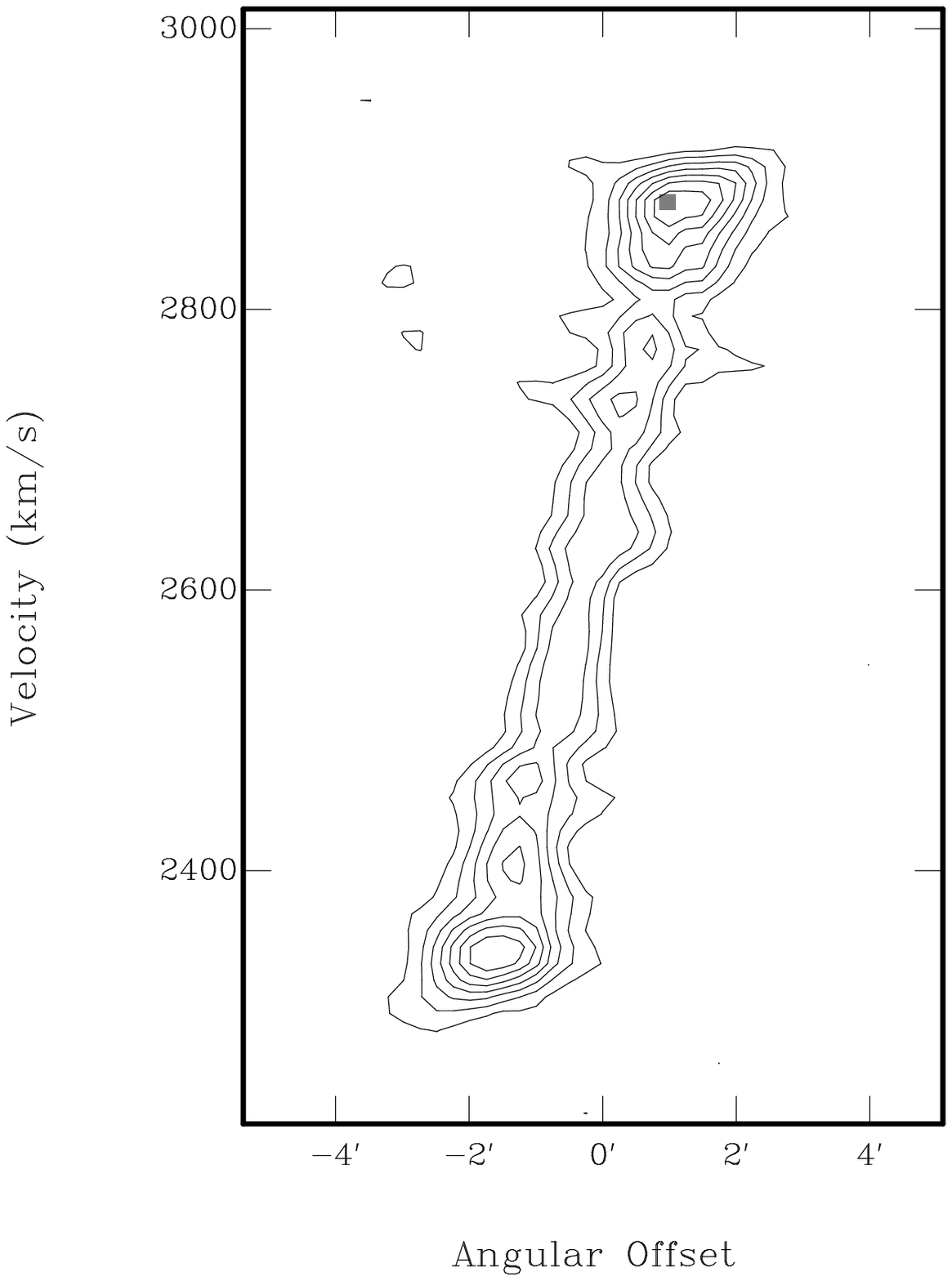,width=7cm,angle=0}}
\caption{{\sl Left:} total \HI\ (contours) superimposed to an optical image of
NGC~3108. 
{\sl Right:} position-velocity slice taken along the major axis of NGC~3108.}
\end{figure}

\subsection {NGC~1947}

NGC 1947 is another minor axis dust--lane elliptical galaxy, that also shows a
ring of CO emission (Sage \& Galletta 1993) and ionized gas (M\"ollenhoff 1982)
with the same position angle as the dust lane.  The stellar component is
rotating around the galaxy's minor axis, perpendicular to the ionized gas
rotation axis (Bertola et al.\ 1992).  We find about $3.4 \times 10^8 M_\odot$
of \HI\ which appears to be in a ring (rotating in the same sense as the CO and
aligned with the dust band) and extended $\sim 10$ kpc each side.  We derive a
value of $M_{\rm HI}/L_B = 0.022$.  Assuming circular orbits and a spherical
mass distribution, we find the $M/L_B$ to be about 8 at $\sim$3 $R_{\rm e}$,
although this value is very uncertain.

\subsection  {IC~5063}

IC~5063 (PKS~2048$-$572) is an early--type galaxy hosting a Seyfert 2 nucleus
and with a dust lane along the major axis.  Its \HI\ content is very high:
$8.4 \times 10^{9} M_\odot$, so that $M_{\rm HI}/L_{B}$ = 0.18.  The total
intensity image shows the \HI\ emission (see Fig.~3) elongated in the
direction of the dust lane ($\sim 38$ kpc each side) and the ionized gas
(Morganti et al.\ 1998).  The \HI\ is mainly distributed in a disk.  Assuming
circular orbits and a spherical mass distribution, we find $M/L_B$ to be
$\sim$ 14 at about 5.4 $R_{\rm e}$.  Strong {\sl very broad} (FWHM of
$\sim$700 \kms) and {\sl blue-shifted} (with respect to the systemic velocity)
absorption against the central radio continuum source is visible in the
position-velocity diagram (Morganti et al.\ 1998, Oosterloo et al.\ 1998). 
Thus, the \HI\ absorption represents the effect of gas outflow resulting from
the interaction of the radio plasma with the ISM gas. 

\begin{figure}
\centerline{\psfig{figure=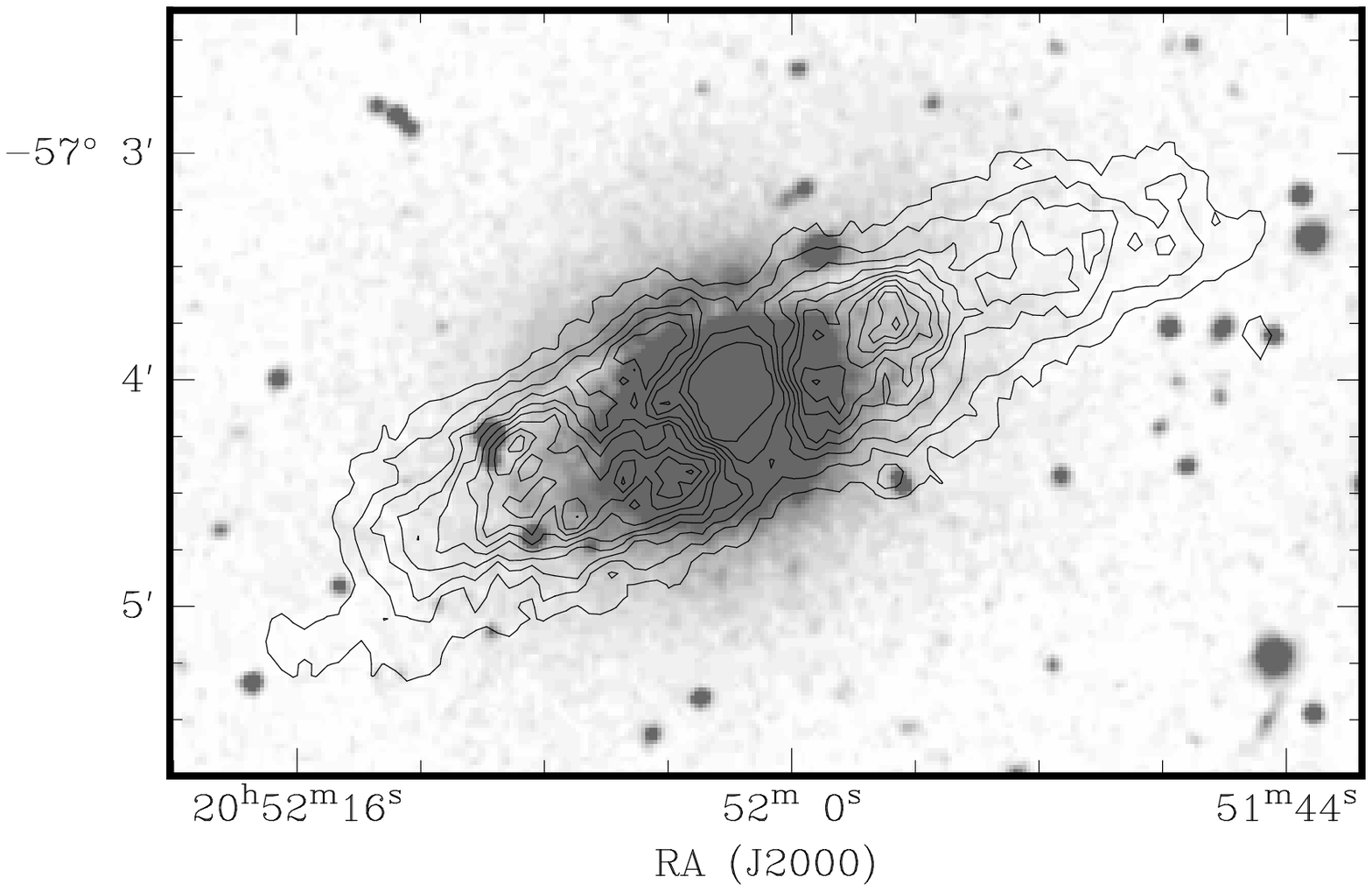,width=8cm,angle=0}
            \psfig{figure=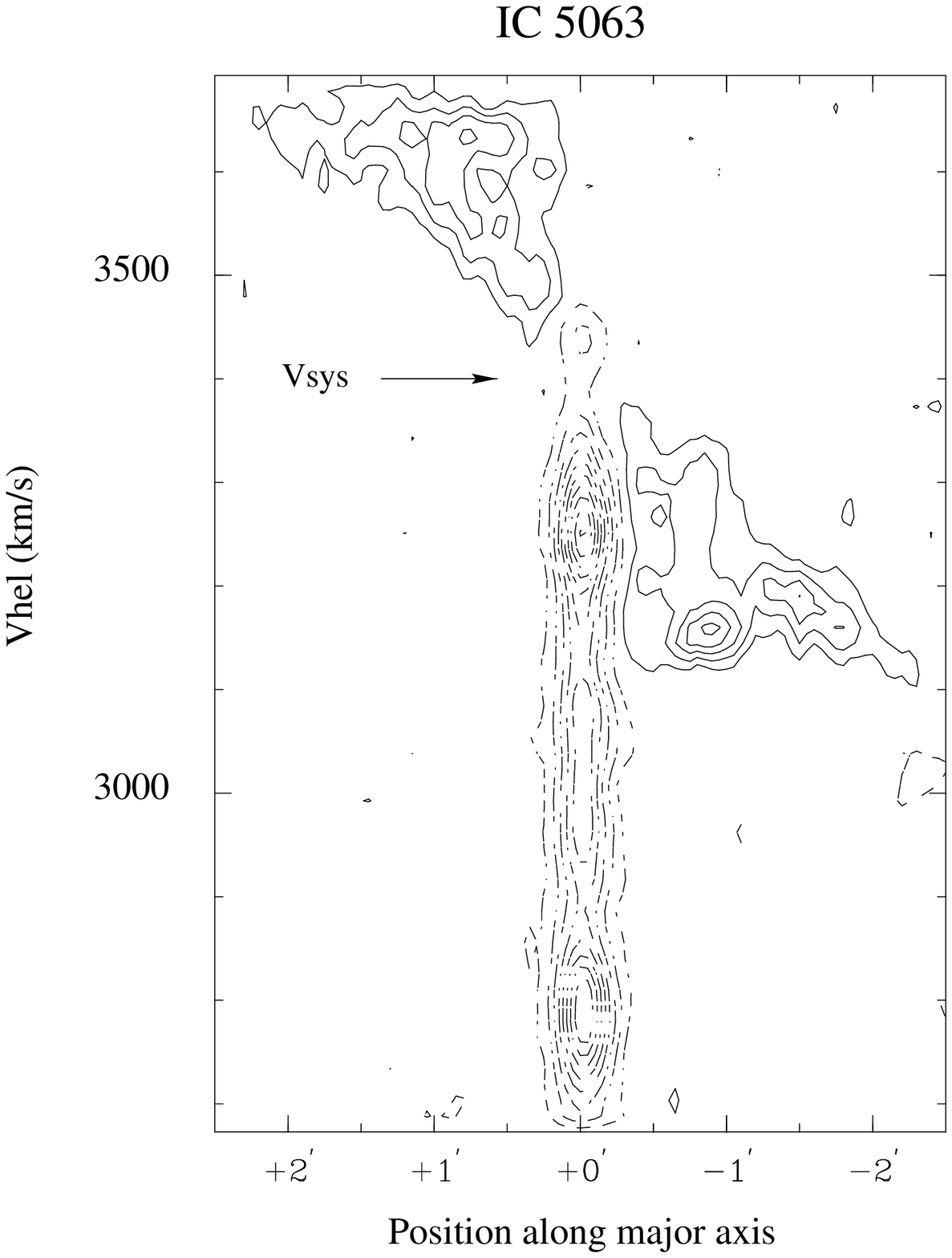,width=8cm,angle=0}}
\caption{{\sl Left:} Total \HI intensity map of IC5063 (contours)
superimposed on the
optical image of the galaxy. {\sl Right}: position-velocity slice taken along
the major axis of IC5063 ({\sl dashed lines} represent the absorption).}
\end{figure}

\section{Discussion}

\subsection {Morphology \& origin of the gas}

Despite being elliptical galaxies, we have detected in three of the four
objects presented {\sl very} large amounts of neutral hydrogen with values of
$M_{\rm \HI}/L_B$ that are at the end of the distribution characteristic of
ellipticals (Knapp 1985) and similar to those typically found in spiral
galaxies.  In the fourth galaxy (NGC 1947) we do detect \HI, but the galaxy is
less hydrogen rich.  Moreover, the \HI\ detected is regularly distributed in
disks or rings, with the possible exception of the outer region in NGC~5266. 
The neutral gas is at much larger radii than are the stars and appears to be
rotationally supported.  The \HI\ is mainly along the dust-lane with the
exception of NGC~5266. 

In the case of NGC~5266, the large amount of neutral gas ($M_{\rm HI}/L_B
\sim$ 0.2) and the \HI\ morphology, suggest that this object could represent
the result of a {\sl major merger}, i.e.\ a merger of two spiral galaxies,
more than 1 Gyr after the interaction of the two spirals (Morganti et al.\
1997b).  If so, NGC~5266 probably represents a relatively old merger remnant
since most of the \HI\ gas appears settled. 
The idea that elliptical galaxies can be the result of a merger between two
disk galaxies has been around for long time (after the seminal work of Toomre
\& Toomre 1972): the pre-formed pairs of disk galaxies can loose their mutual
orbital energy and angular momentum by transferring them from the orbiting
galaxies to the tidal features and then loose also their identities in the
process and blend in an elliptical galaxy.  More recent studies have shown that
the tails formed during the interaction and merger will remain bound to the
remnant (Barnes \& Hernquist 1996 and references therein).  The gas in the
tails will slowly ``fall back'' to the main body of the galaxy in a long time
(for NGC7252 Hibbard \& Mihos (1995) predict that about $10^9 M_\odot$ will fall
within 15-45 kpc in about 3 Gyr). 

The high values of $M_{\rm HI}/L_B$ and the possible luminous counterpart of
the extended neutral gas disks (see below) seems to indicate that other two
objects in our sample (NGC~3108 and IC~5063) could have a similar origin but
would represent a later stage of the merger process.  NGC~5266 is believed to
be well past the phase of ultraluminous IRAS galaxy that should follow the
merger of two spirals.  NGC~3108 would be even more evolved phase (by looking
at the morphology and kinematics of \HI\ and the other IR indicators).  These
systems should be in their {\sl second infall} phase because the neutral gas
appears to be in a quiet phase and their star formation is low (see below):
the gas is falling back and slowly settling. 

The case of NGC~1947 is less clear and indeed the amount of gas and the 
$M_{\rm \HI}/L_B$ seems to indicate that the \HI\ could derive just from 
a small accretion. 

\subsection {The disks}
 
The \HI\ appears to be mainly distributed in disk--like structures.  The \HI\
disks have a low surface density (as shown in Fig.\ 1 for NGC~5266) that falls
well below the threshold needed for star formation to occur.  Thus we would not 
expect widespread star formation to occur, even though the total amount of 
gas is large.  But is there {\sl any} star formation in the \HI\ disks of 
these galaxies?  If we plot the four galaxies on the FIR--radio correlation, 
we find that all (except IC~5063, where the radio continuum emission includes 
a strong contribution from the AGN) follow very well the correlation found for
spirals by Helou et al.\ (1985). The radio continuum is unresolved in our
galaxies (for NGC~3108 only an upper limit has been derived), suggesting that star formation is confined to the central regions.

All four objects have ionized gas in the central region in the form of
disks.  From the available rotation velocity of the ionized gas, these disks
are likely to be inner counterparts of the neutral gas disks.  With the
exception of IC 5063, these objects do not have an AGN and therefore an AGN
cannot account for the ionization of this gas. 

Do we see optical counterparts of the \HI\ disks? In NGC~5266 there is a very
faint optical counterpart of the large scale \HI\ disk (see Fig.\ 1a).  In
NGC~3108, Caldwell (1984) finds evidence of an optical disk.  In IC 5063, the
deep image in Danziger et al.\ (1981) shows a faint large scale structure
(although this will have to be compared with more sensitive low--resolution 
\HI\ images). 

What are these disks and what do they tell us about the structure and the
formation of these objects? The optical counterpart of the disks seen in \HI\
is usually {\sl very} faint, and so it is impossible to verify that these 
stars are 
rotating in a similar way as the gas.  Indeed, these very faint counterparts of
the \HI\ disk could be formed by stars coming from the progenitors and
therefore represent a strong indication that the origin of the gas in these
objects is from a major merger. 

In the case of NGC~1947, no extended low--surface brightness structure is seen
in the optical (Malin priv.\ comm.) and therefore the \HI is more likely
connected to the accretion of a small companion.

\subsection {Dark matter}
 
The objects we have observed show quite a regular distribution of \HI\ and
therefore they are ideal for studying the dark matter properties of
ellipticals, contrary to other objects in our sample (e.g.\ the low--luminosity
ellipticals, Oosterloo et al.\ these proceedings) where we are limited by
resolution for a proper estimate of the $M/L_B$.  With the four dust--lane
ellipticals presented here we find, on average, \HI\ extended up to 5 $R_{\rm
e}$ and values of $M/L_B$ up to 20.  These values are summarized in Fig.\ 4
(modified from Bertola et al.\ 1993).  The plot includes our objects and the
other ellipticals known to have regular and extended \HI.  This gives already
an idea about how few ellipticals have the distribution of neutral gas suitable
for such a study. 

\begin{figure} 
\centerline{\psfig{figure=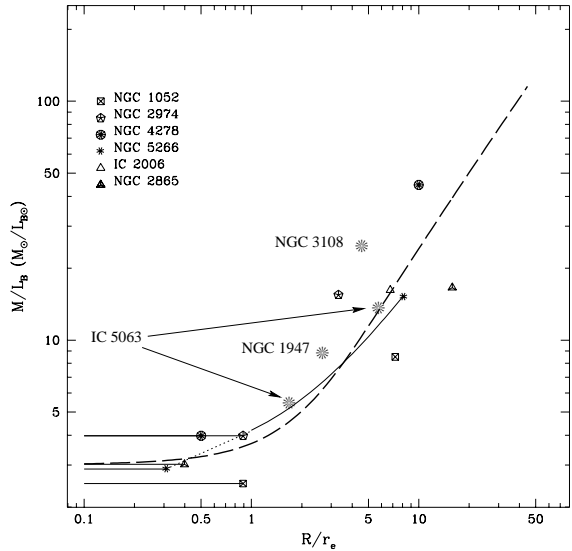,width=8cm,angle=0}} 
\caption{ The log($M/L_B$) - log($R/R_{\rm e}$) 
diagram (modified from Bertola et
al.\ 1993).  The dashed thick line represents the cumulative $M/L_B$ as
function of radius for spiral galaxies.  The symbols represent the data for the
ellipticals obtained from the optical and the \HI\ data.}
\end{figure}

Evidence for flat rotation curves have been found in the \HI\ data in some of
the objects we have observed.  A flat rotation curve is observed in, e.g., 
IC~5063.  Also in the case of NGC~3108, the rotation curve appears to stay (at
least) flat up to 5 $R_{\rm e}$.  Thus, already from the \HI\ data there is
evidence that the dark matter contribution increases at larger radii. 

Fig.\ 4 also includes (for some objects) the values of $M/L$ relative to the
inner parts ($<$1 $R_{\rm e}$) derived from optical data (ionized gas).  By
adding these values the increasing dark matter contribution between the inner
and the outer radii is even more obvious. The trend of $M/L_B$ for spirals as
function of the radius (scaled to the effective radius) is also shown as a 
dashed line.  The values derived for the ellipticals follow the trend found 
for luminous  spirals. 

The values of $M/L_B$ obtained from the \HI\ data for our objects are
comparable with those derived from X--ray observations (Loewenstein these
Proceedings).

\subsection {HI content and radio activity}
 
The idea that clouds of neutral gas from a disk are accreted into the center
and can fuel the nuclear engine was proposed by Gunn 1979.  In our objects we
find that, despite the large amount of neutral gas observed, only one of them
contains an AGN.  This confirms the result of Knapp (1987), that the radio
power does not appear proportional to the \HI\ content.  The activity is likely
to be more connected to the presence of nuclear dust lanes as shown by van
Dokkum \& Franx (1995) in their {\sl HST} study.  However, the similarity in
the optical appearance of some of the systems presented here with, e.g.,
Centaurus~A, rises the question of why they are not radio galaxies like this
object. 

Keeping in mind that the phase of radio activity is relatively short ($\sim
10^7 - 10^8$ yr) compared to the lifetime of a galaxy and the dynamical time
of a merger, it is, therefore, likely that the AGN phase either still has to
come or it is already passed.  Mergers of disk galaxies seem to be quite
efficient in driving gas to the central regions (Barnes \& Hernquist 1991) and
this may favor the idea that the AGN phase already passed.

\section {Conclusions }

There is now a growing number of elliptical galaxies detected and studied in
detail in \HI.  This should soon provide a large enough database to allow us 
to investigate the formation history of these systems in more detail. 

We have presented our results for four dust--lane ellipticals, three of which  have a very large amount of neutral
hydrogen with values of $M_{\rm \HI}/L_B$ that are at the high end of the
distribution characteristic of ellipticals (Knapp 1985) and similar to what
is typically found in spiral galaxies.  Moreover, the \HI\ detected is (often)
regularly distributed in disks or rings. 

To distinguish whether the gas detected is the result of a major merger or
accretion it is not easy.  Van Gorkom \& Schiminovic (1997) pointed out that
there can be two possible ways: from the kinematics or from the amount of \HI. 
Three of the four objects presented here appear to have a {\sl very faint}
optical counterpart of their \HI\ disks where the stars could be (although it
will be very difficult to check) coming from the progenitors.  The same three
objects have more than $10^9 M_\odot$ of \HI\ and values of $M_{HI}/L_B$
similar to what typically found in spirals.  It is hard to see how this could
result from the accretion of just a small companion.  Thus, these galaxies may
represent the results of major mergers as seen at a late stage of the
evolution.  In these objects the secondary infall of gas has produced regular
low surface density disks where no star formation is happening (with the
possible exception of the very central region).  In the remaining galaxy, NGC
1947, the dust lane is probably connected to the accretion of a small
companion. 

The regular \HI\ distribution and kinematics allows us has to derive values of
$M/L_B \sim 20$ at distances of $\sim$5 $R_{\rm e}$ from the centre.  Evidence
for an increase of the dark matter contribution at larger radii can also be
derived from the flat rotation curves observed in some of the observed
galaxies.

\bigskip

{\bf Questions}

\smallskip

{\sl Bruzual:} Do you have any information about the color of the stellar
populations in the central part and the extended disks of NGC~5266.  It would
be very interesting to know the color difference in order to date the encounter
from the stellar evolution point of view. 

\smallskip

{\sl Morganti:} Unfortunatly we do not have information on the color of the
extended optical halo and my guess is that it will also be very difficut to
obtain this information because this halo is very faint. 

\smallskip

{\sl Young:} Of the seven dust lane ellipticals you observe, three were not
detected. Do you know why they were not detected? Might they have smaller
gas/dust ratios than the four detected ones?

\smallskip

{\sl Morganti:} From a preliminary look at their characteristics ($M_{\rm
\HI}/L$ etc.), we did not find any obvious difference between the objects
detected and those undetected in \HI.


\begin{references}

\reference Barnes J.E. \& Hernquist L.E. 1991, ApJ 307, 65

\reference Barnes J.E. \& Hernquist L.E. 1996, ApJ 471, 115

\reference Bertola F., Pizzella A., Persic M.  \& Salucci P.  1993, \apj, 416,
L45

\reference Bertola F., Galletta G., Zeilinger W.W. 1992, A\&A 254, 89

\reference Bregman J.N., Hogg D.E. \& Roberts M.S. 1992, \apj, 387, 484

\reference Caldwell, N. 1984, \apj, 278, 96. 

\reference Danziger J.I., Goss W.M., Wellington K.J. 1981, \mnras, 196, 845

\reference Franx, M., van Gorkom, J.H. and de Zeeuw, T., 1994, ApJ 436, 642  

\reference Gunn J.E. 1979, in {\sl ``Active Galactic Nuclei''}, C.Hazard \&
S.Mitton eds. (Cambridge Univ.) p.213

\reference Helou G., Soifer B.T. \& Rowan-Robinson M. 1985, \apj, 298, L7

\reference Hibbard J.E. \& Mihos J.C. 1995, \aj, 110, 140  

\reference Hibbard J.E. \& van Gorkom 1996, AJ 111, 655

\reference Kotanyi C. \& Ekers R.D. 1979, A\&A, 73, L1 

\reference Knapp G.R., Turner E.L. \& Cunniffe P.E. 1985, \aj, 90, 454 

\reference Knapp G.R. 1987, in {\sl Structure and dynamics of elliptical
galaxies} IAU Symp. 127, p. 145

\reference Lees J.F. 1994, in {\sl Mass-Transfer Induced Activity in
Galaxies''}, ed.  Shlosman I., Cambridge Uni.  Press, p.  432

\reference M\"ollenhoff C. 1982,  A\& A, 108, 130

\reference M\"ollenhoff C., Hummel E. \& Bender R. 1992, A\&A, 255, 35

\reference Morganti R., Oosterloo T., Sadler E.M.  1997a, in {\sl ``The Nature
of Elliptical Galaxies''}, M.Arnaboldi, G.S.  Da Costa \& P.Saha, eds, ASP
Conf.Series p.349
 
\reference Morganti R., Sadler M.E., Oosterloo T.A., Pizzella A., Bertola F.
1997b, \aj, 113, 937

\reference Morganti R., Oosterloo T., Tsvetanov Z., 1998, AJ 115, 915

\reference Oosterloo T., Morganti R., Tzioumis A., Reynolds J.  1998, in {\sl
Radio Emission from Galactic and Extragalactic Compact Sources}, ASP
Conference Series, Volume 144, eds.  J.A.  Zensus, G.B. 
Taylor,  J.M.  Wrobel, p.  197. 

\reference Raimond E. et al. 1981, \apj, 246, 708

\reference Sadler E.M. \& Gerhard O.E. 1985, \mnras, 214, 177

\reference Sage L.\& Galletta G. 1993, \apj, 419, 544 

\reference Schweizer F.  1995, in {\sl ``Stellar Populations''}, van der Kruit
\& Gilmore, eds, IAU Symp.  164, Kluwer, p.275 \reference Toomre A.  \& Toomre
J.  1972, \apj, 178, 623

\reference van Gorkom et al.  1986, \aj, 91, 791 

\reference van Gorkom J.H.  \&
Schiminovic D.  1997, in {\sl ``The Nature of Elliptical Galaxies''}
M.Arnaboldi, G.S.  Da Costa \& P.Saha, eds, ASP Conf.Series, p. 310

\reference van Gorkom J.H.  1992, in {\sl ``Morphological and Physical
Classification of Galaxies''}, Longo G., Capaccioli M., Busarello G., eds,
Kluwer Academic Publisher p.  233

%\reference Mihos J.C. \& Hernquist L.E. 1994, \apj, 431, L9
%Schiminovich D., van Gorkom J.H., van der Hulst J.M. \& Malin D.F., 1995, \apj,
%444, L77 
%Schiminovich D., van Gorkom J.H., van der Hulst J.M. \& Kasow S., 1994, \apj,
%423, L101 
%\reference Phillips M.M., Jenkins C.R., Dopita M.A., Sadler E.M. \& Binette L., 1986, \aj,
%91, 1062
%Sadler E.M., 1984, \aj, 89, 53 
%\reference van Gorkom J.H., Knapp G.R., Ekers R.D., Ekers D.D., Laing R.A.  \&
%Polk K.  1989, \aj, 97, 708
%\reference Schweizer F., van Gorkom J.H. \&  Seitzer P. 1989, \apj, 338, 770 


\end{references}
\end{document}